# First-principles study of electronic band structure and elastic properties of superconducting nanolaminate Ti$_2$InC.

I.R. Shein,* A.L. Ivanovskii

*Institute of Solid State Chemistry of the Ural Division of the Russian Academy of Sciences, 620041, GSP-145, Ekaterinburg, Russia*

**Abstract**

The full-potential linearized augmented plane wave method with the generalized gradient approximation for the exchange-correlation potential (FLAPW-GGA) is used to predict the electronic and elastic properties of the newly discovered superconducting nanolaminate Ti$_2$InC. The band structure, density of states and Fermi surface features are discussed. The optimized lattice parameters, independent elastic constants, bulk and shear moduli, compressibility are evaluated and discussed. The elastic parameters of the polycrystalline Ti$_2$InC ceramics are estimated numerically for the first time.

___________________________________________
* Corresponding author
  E-mail shein@ihim.uran.ru,
  Phone: +7 343 362 33 51, Fax: +7 343 374 44 95

## 1. Introduction

The layered ternary carbides and nitrides with the general formula $M_{n+1}A(C,N)_n$ (the so-called *nanolaminates* [1], where M are *d* transition metals, A - *p* elements belonging to the groups III - VI, and n = 1, 2 and 3) have gained recently considerable attention as "functional ceramics" owing to a unique combination of metallic and ceramic properties including high strength and stiffness at high temperatures, resistance to oxidation and thermal shock, as well as high Young's moduli, low hardness and ultra-low solid friction coefficients, see [1-8].

Furthermore, for some nanolaminates (of the general formula $M_2AC$, also known as ''MAX'' [9], "H" or "211" phases), low-temperature superconductivity was discovered. Until recently $Mo_2GaC$ ($T_C$ ~ 3.7 ÷ 4.1K) [10], $Nb_2SC$ ($T_C$ < 5K) [11], $Nb_2SnC$ ($T_C$ ~ 7.8K) [12], and $Nb_2AsC$ ($T_C$ ~ 2K) [13] superconductors were known; for some of them the features of their electronic band structure were elucidated using the density functional calculations [13-15].

Quite recently, superconductivity with the critical temperature $T_C$ ~ 3.1K was reported for the nanolaminate $Ti_2InC$ [16]. This material belongs to the family of the mentioned H phases and adopts the $Cr_2AlC$ –type hexagonal structure (space group $P6_3/mmc$) which consists of rocksalt-like TiC layers separated by atomic In sheets.

In the present work, we have performed first-principles calculations of the newly discovered superconducting nanolaminate $Ti_2InC$ in order to predict its elastic and electronic properties. As a result, the optimized lattice parameters, independent elastic constants, bulk and shear moduli, as well as the electronic band structure, density of states and Fermi surface are evaluated and discussed. In addition, numerical estimates of the elastic parameters of the polycrystalline Ti$_2$InC ceramics are performed for the first time.

## 2. Computational method

The Ti$_2$InC nanolaminate crystallizes in the Cr$_2$AlC - type crystal structure, with the unit cell containing two formula units, and the atoms occupy the Wyckoff positions: Ti: 4*f* {(1/3, 2/3, *z*), (2/3, 1/3, *z*+1/2), (2/3, 1/3,–*z*), (1/3, 2/3,–*z*+1/2)}; In: 2*d* {(1/3, 2/3, 3/4), (2/3, 1/3, 1/4)} and carbon: 2*a* {(0, 0, 0), (0, 0, 1/2)}, where *z* is the so-called internal parameter, see [1,9].

Our calculations were carried out by means of the full-potential method with the mixed basis APW+lo (FLAPW) implemented in the WIEN2k suite of programs [17]. The generalized gradient approximation (GGA) to exchange-correlation potential in the PBE form [18] was used. The plane-wave expansion was taken to $R_{MT} \times K_{MAX}$ equal to 7, and the *k* sampling with 10×10×4 *k*-points in the Brillouin zone was used. The calculations were performed with full-lattice optimizations; the self-consistent calculations were considered to be converged when the difference in the total energy of the crystal did not exceed 0.01 mRy as calculated at consecutive steps.

## 3. Results and discussion.

### *3.1. Structural and elastic properties.*

Firstly, the equilibrium lattice parameters for the nanolaminate Ti$_2$InC were calculated. These values ($a^{calc}$ =3.1373 Å, $c^{calc}$ =14.1842 Å and the internal parameter *z* = 0.0783) are in reasonable agreement with the available experiment [16]: the divergences (*($a^{calc}$ - $a^{exp}$)/$a^{exp}$* = 0.0046 and *($c^{calc}$ - $c^{exp}$)/$c^{exp}$* = 0.0088) between the calculated and experimentally observed parameters should be attributed to the well known overestimation of structural parameters for GGA calculations.

Secondly, the values of five independent elastic constants ($C_{ij}$, namely $C_{11}$ =273.4 GPa, $C_{12}$ = 62.9 GPa, $C_{13}$ =50.3 GPa, $C_{33}$= 232.3 GPa and $C_{44}$ =87.2 GPa, while $C_{66}$ = ½ ($C_{11}$ - $C_{12}$)) were evaluated by calculating the stress tensors on different deformations applied to the equilibrium lattice of the hexagonal unit cell, whereupon the dependence between the resulting energy change and the deformation was determined; see [19-21] for details. All these elastic constants are positive and satisfy the well-known Born's criteria for mechanically stable hexagonal crystals: $C_{11}$ >0, ($C_{11}$ - $C_{12}$) >0, $C_{44}$ >0, and ($C_{11}$ + $C_{12}$)$C_{33}$ - 2$C_{12}^2$ > 0.

The calculated elastic constants allowed us to obtain the macroscopic mechanical parameters of Ti$_2$InC, namely its bulk (*B*) and shear (*G*) moduli, which were calculated using the Voigt (V) [22] and Reuss (R) [23] approximations. These values were: $B_V$ = 122.9 GPa, $B_R$ = 117.0 GPa and $G_V$ = 97.0 GPa, $G_R$ =96.2 GPa.

The above elastic parameters were obtained from first-principle calculations of the Ti$_2$InC monocrystal. Meanwhile, the majority of the synthesized and experimentally examined samples of the nanolaminate Ti$_2$InC were prepared and investigated as polycrystalline ceramics [16,24-26], *i.e.* in the form of aggregated mixtures of microcrystallites with a random orientation. Thus, it is useful to estimate the elastic parameters for the polycrystalline Ti$_2$InC ceramics.

For this purpose we utilized the Voigt-Reuss-Hill (VRH) approximation. In this approach, according to Hill [27], the two above mentioned approximations (Voigt and Reuss) were used. In this way, when the bulk modulus ($B_{VRH}$) and the shear modulus ($G_{VRH}$) were obtained from $B_{V,R}$ and $G_{V,R}$ using the VRH approach in simple forms as: $B_{VRH}$ = ½ ($B_V$ + $B_R$) and $G_{VRH}$ = ½ ($G_V$ + $G_R$), the averaged compressibility ($\beta_{VRH}$ = 1/$B_{VRH}$) and Young's modulus (Y$_{VRH}$) were calculated from the expression:

$$Y_{VRH} = \frac{9B_{VRH}G_{VRH}}{3B_{VRH}+G_{VRH}}$$

Then, the Poisson's ratio (ν) was obtained for the polycrystalline Ti$_2$InC from B$_{VRH}$, G$_{VRH}$ and Y$_{VRH}$ as:

$$v = \frac{3B_{VRH} - 2G_{VRH}}{2(3B_{VRH} + G_{VRH})}$$

Here we should point out that our estimations were made in the limit of zero porosity of Ti$_2$InC ceramics. The above-mentioned parameters listed in Table 1 allowed us to make the following predictions.

(i). From our results we see that for Ti$_2$InC $B_{VRH} > G_{VRH}$; this implies that a parameter limiting the mechanical stability of this material is the shear modulus $G_{VRH}$.

(ii). According to the criterion [35], a material is brittle if the *B/G* ratio is less than 1.75. In our case $B/G \sim 1.3$, therefore these materials will behave in a brittle manner.

(iii). The bulk, shear and Young's moduli for Ti$_2$InC are rather small as compared with other related nanolaminates Ti$_2$AC (where A are p elements of III-VI groups) and are much smaller than for the cubic TiC (see Table 1), whereas the compressibility of Ti$_2$InC ($\beta_{VRH} \sim 0.00834$ 1/GPa) adopts the maximal value. Thus, as compared with other Ti$_2$AC phases, Ti$_2$InC is a relatively soft material.

(iv). The values of the Poisson ratio (*v*) for covalent materials are small ($v = 0.1$), whereas for ionic materials a typical value of v is 0.25 [36]. In our case, the value of *v* for Ti$_2$InC is about 0.19, *i.e.* a considerable ionic contribution in intra-atomic bonding should be assumed for this nanolaminate. Besides, for covalent and ionic materials, the typical relations between bulk and shear moduli are $G \approx 1.1B$ and $G \approx 0.6B$, respectively. In our case, the calculated value of $G_{VRH}/B_{VRH}$ is 0.80, which also indicates that the mixed ionic-covalent bonding is suitable for Ti$_2$InC.

(v). To evaluate the elastic anisotropy of Ti$_2$InC, the so-called anisotropic factor $A = C_{33}/C_{11}$ was calculated. Here the value $A = 1$ represents completely elastic isotropy, while values smaller or greater than 1 point to the degree of elastic anisotropy. Other estimations follow from anisotropy factors of compression ($A_{comp} = \{(B_V - B_R)/(B_V + B_R)\} \cdot 100\%$) and shear ($A_{shear} = \{(G_V -$ $G_R)/(G_V + G_R)\}\cdot 100\%$) [37]. For the isotropic crystals, the values of $A_{comp}$ and $A_{shear}$ are equal to zero, whereas the value 100% means the maximal anisotropy. Our calculations show that $C_{33}/C_{11} = 0.85$, $A_{comp} = 2.4\%$ and $A_{shear} = 1.1\%$. This means that the nanolaminate $Ti_2InC$ will exhibit small elastic anisotropy.

*3.2. Electronic properties.*

Figure 1 displays the calculated band structure of $Ti_2InC$ along some high-symmetry directions of the Brillouin zone (BZ) in the energy range from -15 eV to +5 eV. In the regions about -14 eV and -11÷ -10 eV below the Fermi level ($E_F$), the low-dispersive In $4d$ and C $2s$-like bands are located, whereas occupied valence bands are in the interval from -8.6 eV to $E_F$. The valence and conduction bands overlap considerably and there is no band gap at the Fermi level, which is indicative of metallicity of this material. It is also seen that three bands intersect the Fermi level (one band around K to M, and two bands around A to M), which points out to a multi-band character of this system. An interesting feature is a 2D-like behavior of quasi-flat electronic bands along the $c$ axis (from L-M and from K-H). It shows that the interactions between the adjacent layers in the nanolaminate $Ti_2InC$ are relatively small.

The corresponding Fermi surface (FS) in the first BZ is shown in Fig. 2. Owing to the quasi-two-dimensional electronic structure, the Fermi surface consists of two main sheets centered along the Γ-A high symmetry line, *i.e.* parallel to the $k_z$ direction. The first sheet is cylindrical-like, and the second sheet adopts a prismatic-like form. Finally, the 3D ellipsoid-like sheets are centered at the K point.

As electrons near the Fermi surface are involved in the formation of the superconducting state, it is important to understand their nature. The total and atomic and orbital decomposed partial DOSs are shown in Fig. 3. It is seen that the Fermi level lies in a local DOS minimum, reflecting the separation between the bonding and antibonding states. The total DOS at the Fermi level ($N(E_F)$ =1.223 states/(eV·form.unit)) is formed mainly by Ti $3d$ states (0.971 states

/(eV·form.unit)), while the admixtures of In and carbon states are much smaller (about 0.001 ÷ 0.035 states /(eV·form.unit)). The same feature, i.e. the dominant role of transition metal $d$ states in the near-Fermi region, was found also for other superconducting nanolaminates [13-15]. Finally, the obtained data also allow us to estimate the Sommerfeld constant ($\gamma$) for Ti$_2$InC assuming the free electron model $\gamma = (\pi^2/3)N(E_F)k^2_B$. The calculated value is $\gamma = 2.883$ mJ·K$^{-2}$·mol$^{-1}$.

## 4. Conclusion

Employing the first principles FLAPW-GGA approach, we have studied the elastic and electronic properties of the newly discovered superconducting nanolaminate Ti$_2$InC.

Our analysis shows that the nanolaminate Ti$_2$InC is a mechanically stable material with small elastic anisotropy; the parameter limiting its mechanical stability is the shear modulus. In addition, Ti$_2$InC is a relatively soft material with high compressibility and will behave in a brittle manner.

The band structure calculations reveal a 2D-like behavior of the quasi-flat electronic bands along the $c$ axis. Therefore the interactions between the adjacent layers in the nanolaminate Ti$_2$InC are rather small.

The near-Fermi bands involved in the formation of the superconducting state are formed mainly by Ti 3$d$ states, while the admixtures of valence states of indium and carbon sublattices are much smaller.

**Acknowledgement**

Financial support of the RFBR (Grant 07-03-96061) is gratefully acknowledged.

# References

[1] M.W. Barsoum, Prog. Solid State Chem. 28 (2000) 201.
[2] A.L. Ivanovskii, Uspekhi Khimii 65 (1996) 499.
[3] Z.M. Sun, R. Ahuja, J.M. Schneider, Phys. Rev. B 68 (2003) 224112.
[4] M.W. Barsoum, T. El-Raghy, J. Am. Ceram. Soc. 79 (1996) 1953.
[5] M.W. Barsoum, D. Brodkin, T. El-Raghy, Scr. Metall. Mater. 36 (1997) 535.
[6] J.-F. Li, W. Pan, F. Sato, R. Watanabe, Acta Mater. 49 (2001) 937.
[7] S. Amini, M.W. Barsoum, T. El-Raghy, J. Am. Ceram. Soc. 90 (2007) 3953.
[8] N.I. Medvedeva, A.J. Freeman, Scripta Mater. 58 (2008) 671.
[9] M.W. Barsoum, Physical properties of the MAX phases, in: Encyclopedia of Materials: Science and Technology, Elsevier, Amsterdam, 2006.
[10] L.E. Toth, W. Jeitschko, M. Yen, J. Less Common Met. 10 (1967) 129.
[11] K. Sakamaki, H. Wada, H.Y. Nozaki, Y. Onuki, M. Kawai. Solid State Commun. 112 (1999) 323.
[12] A.D. Bortolozo, O.H. Sant'Anna, M.S. da Luz, C.A.M. dos Santos, A.S. Pereira, K.S. Trentin, A.J.S. Machado. Solid State Commun. 139 (2006) 57.
[13]. S.E. Lofland, J.D. Hettinger, T. Meehan, A. Bryan, P. Finkel, S. Gupta, M. W. Barsoum, G. Hug. Phys. Rev. B74 (2006) 174501.
[14] S.V. Halilov, D.J. Singh, D.A. Papaconstantopoulos. Phys. Rev. B 65, 174519 (2002).
[15] I.R. Shein, V.G. Bamburov, A.L. Ivanovskii. Doklady Phys. Chem. 411 (2006) 317.
[16] A.D. Bortolozo, O.H. Sant'Anna, C.A.M. dos Santos, A.J.S. Machado. Solid State Commun. 144 (2007) 419.
[17] P. Blaha, K. Schwarz, G. Madsen et al. WIEN2k, An Augmented Plane Wave Plus Local Orbitals Program for Calculating Crystal Properties, Vienna University of Technology, Vienna, 2001.
[18] J.P. Perdew, S. Burke, M. Ernzerhof, Phys. Rev. Lett. 77 (1996) 3865.
[19] F. Jona, P.M. Marcus, Phys. Rev. B 66 (2002) 094104.
[20] K.B. Panda, K.S.R. Chandran, Comput. Mater. Sci. 35 (2006) 134.
[21] I.R. Shein, A.L. Ivanovskii, J. Phys.: Cond Matter 20 (2008) 415218.
[22] W. Voigt, Lehrbuch der Kristallphysik (Teubner, Leipzig, 1928).
[23] A. Reuss, Z. Angew. Math. Mech. 9 (1929) 49.
[24] M.W. Barsoum, J. Golczewski, H.J. Seifert, F. Aldinger. J. Alloys Comp. 340 (2002) 173.
[25] A. Ganguly, M.W. Barsoum, J. Schuster. J. Am. Chem. Soc. 88 (2005) 1290.
[26] S. Gupta, E.N. Hoffman, M.W. Barsoum. J. Alloys Comp. 426 (2006) 168.
[27] R. Hill, Proc. Phys. Soc. London, A65 (1952) 349.
[28] Y.L. Du, Z.M. Sun, H. Hashimoto, W.B. Tian. Phys. Letters A 372 (2008) 5220.
[29] A. Bouhemadou, R. Khenata. J. Appl. Phys. 102 (2007) 043528.

[30]. J. D. Hettinger, S.E. Lofland, P. Finkel, T. Meehan, J. Palma, K. Harrell, S. Gupta, A. Ganguly, T. El-Raghy, M.W. Barsoum. Phys. Rev. B 72 (2005) 115120.

[31]. A. Bouhemadou, Physica B 403 (2008) 2707.

[32] J.A. Warner, S.K.R. Patil, S.V. Khare, K.C. Masiulaniec. Appl. Phys. Lett. 88 (2006) 101911.

[33] G. Hug. Phys. Rev. B 74 (2006) 184113.

[34] V.V. Ivanovskaya, I.R. Shein, A.L. Ivanovskii. Diamond Related Mater. 16 (2007) 243.

[35] S.F. Pugh, Phil. Mag. 45 (1954) 833.

[36] J. Haines, J.M. Leger, G. Bocquillon, Ann. Rev. Mater. Res. 31 (2001) 1.

[37] D.H. Chung, W.R. Buessem, in: F.W. Vahldiek, S.A. Mersol (Eds.), Anisotropy in Single Crystal Refractory Compounds, Plenum, New York, 1968, pp. 217–245.

Table 1. Calculated bulk modulus ($B$ in GPa), shear modulus ($G$ in GPa), Young's modulus ($Y$, in GPa), and Poisson's ratio ($v$) of $Ti_2InC$ compared with available data for other nanolaminates $Ti_2AC$ and TiC.

| system | B | G | Y | $v$ |
|---|---|---|---|---|
| $Ti_2InC$ | 123/117 (124) ** | 97/95 (96) | (228) | (0.184) |
| $Ti_2SC$ [28] | 181 | 134 | 322 | 0.200 |
| $Ti_2GaC$ [29] | 140 | 130 | 282 | 0.165 |
| $Ti_2AlC$ [30] | 144 | 118 | 277 | 0.19 |
| $Ti_2SnC$ [31] | 159 | 114 | 275 | 0.212 |
| $Ti_2TlC$ [32] | 125 | - | - | - |
| $Ti_2GeC$ [33] | 163 | - | - | - |
| $Ti_2PbC$ [33] | 140 | - | - | - |
| $Ti_2PC$ [33] * | 163 | - | - | - |
| $Ti_2AsC$ [33] * | 175 | - | - | - |
| $Ti_2SiC$ [33] * | 167 | - | - | - |
| TiC [34] | 240 | - | 447 | 0.19 |

* hypothetical phases.

** in Voigt / Reuss approximations and for polycrystalline ceramics - in parentheses.

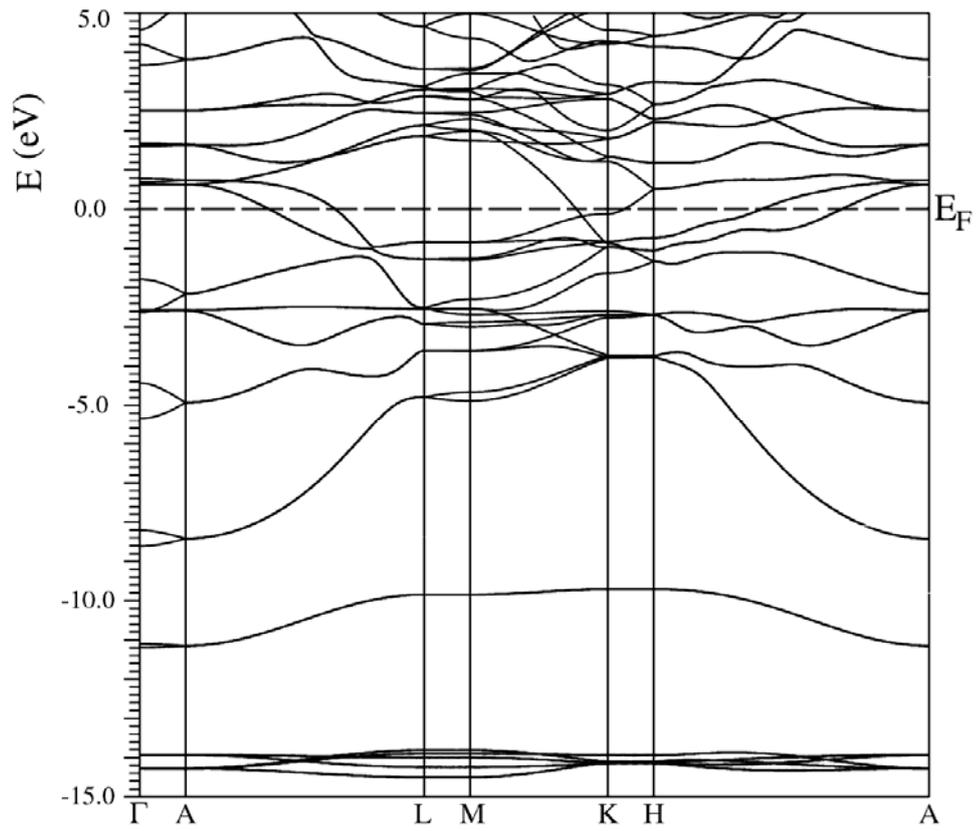

Figure 1. Electronic bands of Ti$_2$InC

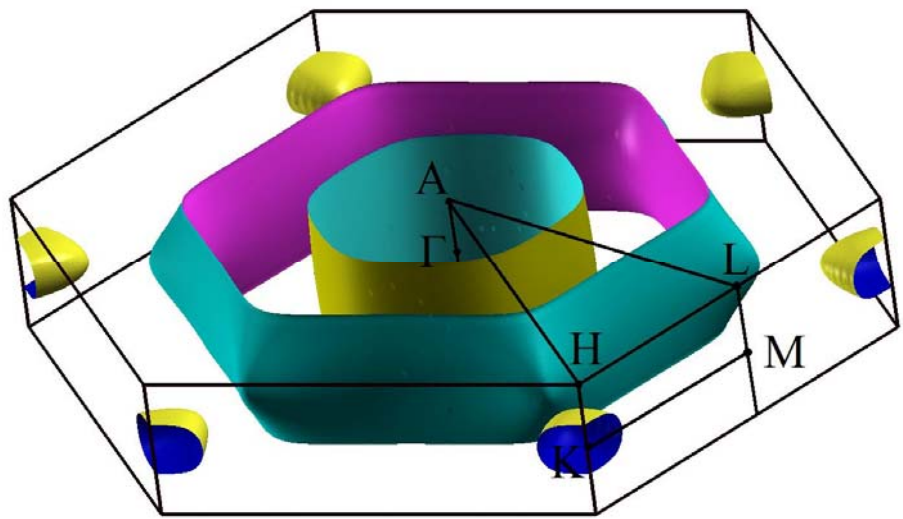

Figure 2. (*Color online*) Fermi surface of Ti$_2$InC

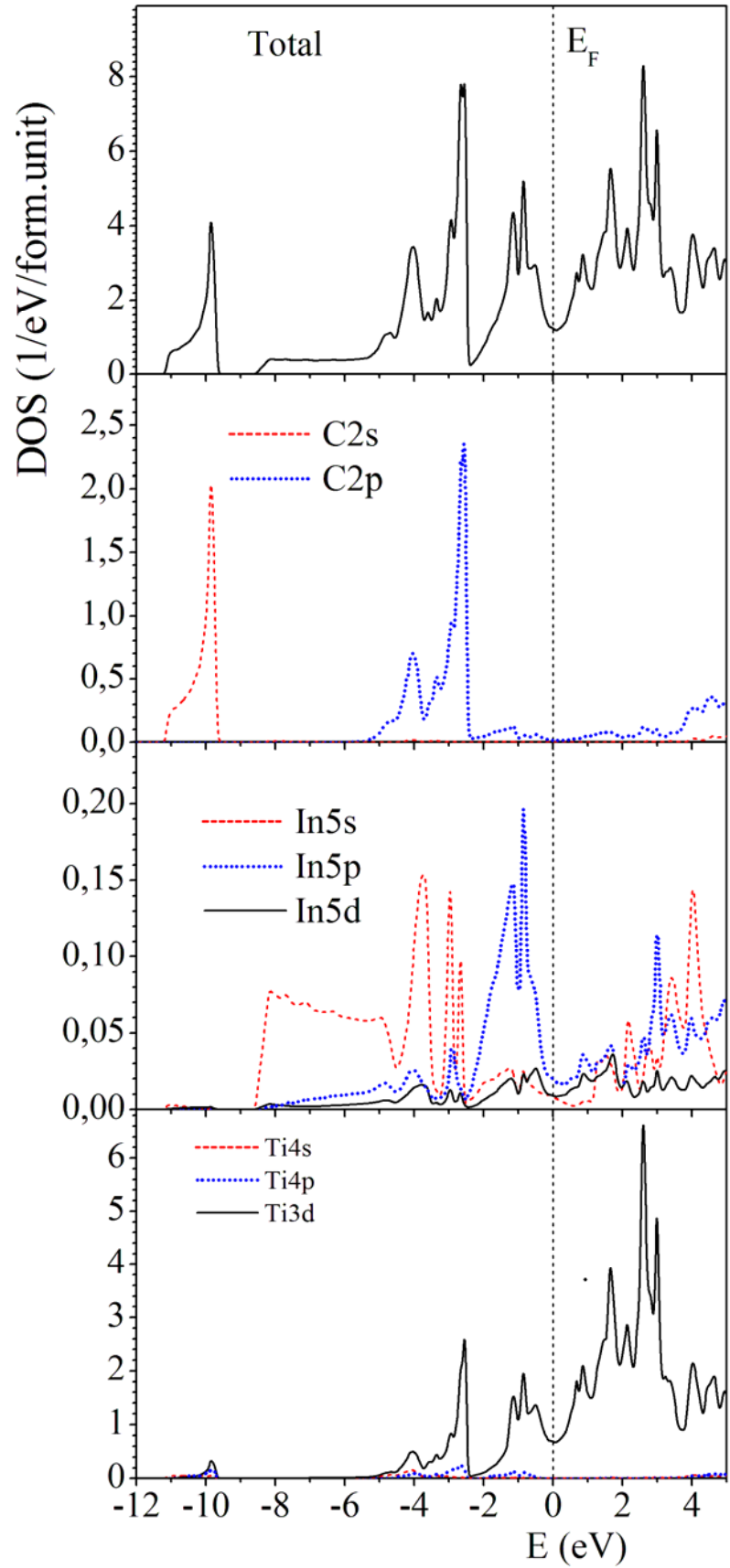

Figure 3. (*Color online*) Total and partial densities of states of Ti$_2$InC.